\documentclass[aps,prb,reprint,superscriptaddress]{revtex4-2} 

\usepackage{graphicx} 
\usepackage{amsmath}  
\usepackage{amssymb}  
\usepackage{dcolumn}  
\usepackage{bm}       
\usepackage{float}    
\usepackage{ragged2e} 
\usepackage{parskip}
\usepackage{soul}

\begin{document}

\title{Unraveling the temperature-responsive charge-disproportionation in BaBiO$_3$}

\author{Sumit Sarkar$^*$}
\affiliation{Department of Physics and Astronomy, University of Waterloo, Waterloo, ON N2L 3G1, Canada.}
\affiliation{UGC-DAE Consortium for Scientific Research, Indore 452001, India.}
\author{Priyanka Yadav}
\affiliation{Department of Physical Sciences, Indian Institute of Science Education and Research (IISER) Mohali,
Knowledge City, Sector 81, Mohali 140306, India }
\author{Sourav Chowdhury}
\affiliation{Deutsches Elektronen-Synchrotron DESY, Notkestrasse 85, 22607, Hamburg, Germany}
\author{Rajamani Raghunathan}
\affiliation{UGC-DAE Consortium for Scientific Research, Indore 452001, India.}
\author{Ram Janay Choudhary$^*$}
\affiliation{UGC-DAE Consortium for Scientific Research, Indore 452001, India.}

\begin{abstract}
\justifying 
This study shows that the charge disproportionation at the Bi site in BaBiO$_3$ alters as a function of temperature. Decreasing the temperature from 300K down to 160K leads to a significant modification of the density of states corresponding to the Bi-O hybridized band near the Fermi level (E$_\text{F}$). This modification indicates reduction of Bi 6$sp$ - O 2$p$ hybridization and O 2$p$ spectral weight near E$_\text{F}$. The strong decrement of covalency at lower temperatures is accompanied by a decrement in O 2$p$ hole density due to possible charge transfer from Bi 6$s$ to the O 2$p$ band. Bi-charge state analysis from Bi-4$f$ core-level spectra showed that at 300K, $\delta$ (charge difference between alternate Bi sites) value in 4$\pm\delta$ is much less than at 160K, which reveals the transition towards the ionic nature of CD or static CD in BBO at low temperature. On the other hand, O 1$s$ core-level spectra displayed an asymmetric shape, and temperature-dependent modifications of the asymmetric shape and intensity have been observed. This highlights the significant influence of the O 2$p$ band hole on the dynamical CD at the Bi site.
\end{abstract}

\maketitle 


\section{Introduction}
In partially filled valence band perovskite systems, the atomic orbital contribution at the Fermi level (E$_\text{F}$) is crucial in stabilizing the electronic ground state \cite{ref1,ref2,ref3,ref4}. According to the band theory, a partially filled band should exhibit a metallic behavior. However, in certain cases, the system remains insulating, despite a half-filled valence band due to the presence of strong correlation between electrons. The type of insulating character observed in these systems is determined by the nature of the low energy inter-orbital charge fluctuation near E$_\text{F}$. Depending on the orbital nature across E$_\text{F}$ can be realized a Mott-Hubbard or a charge-transfer insulator \cite{ref2,ref3,ref5AshcroftMermin,ref6,ref7,ref8}. Also, competition between onsite and long-range Coulomb interaction ($U$ and $V$) has been shown to result in electronic phases like the charge density wave (CDW) or charge-ordered state (CO), spin density wave (SDW), superconductivity etc. \cite{ref9,ref10,ref11,ref12}. According to the Zaanen, Sawatzky, and Allen (ZSA) phase diagram, the competition between (i) intra-atomic Coulomb interaction ($U$), (ii) Metal-ligand site energy difference ($\Delta$), (iii) Metal to ligand hopping interaction strength ($t_{ml}$) and (iv) Inter-atomic Coulomb interaction ($V$) determines the nature of the lowest energy charge fluctuations across the E$_\text{F}$ \cite{ref8}. Based on the relative values of these parameters, transition metal compounds can be classified into two main types of insulating phases in the ZSA phase diagram: 1. Mott-Hubbard (MH) insulators with $U < \Delta$\cite{ref13,ref14,ref15}, and 2. Positive charge-transfer (positive-CT) insulators or metals with $U > \Delta$, in which the charge transfer from the metal to the ligand orbitals is energetically favorable. The charge-fluctuation in MH and positive-CT insulator can be written as $d^n+d^n = d^{n-1}+d^{n+1}$ and $d^n+d^n = d^n \underline{L} + d^{n+1}$ respectively \cite{ref8,ref16,ref17}. It is important to be noted that in the charge-transfer regime, the insulating or metallic behaviour of transition metal compounds is correlated with the energy difference between the 3$d$ orbitals and the oxygen 2$p$ orbital. It states that when this energy difference ($\Delta$) is large, the compound tends to be insulating, whereas for small $\Delta$ values, the compound tends to be metallic, called correlated metal \cite{ref18,ref19}. On the other hand, in certain 3$d$ oxide systems, a distinct type of insulating behaviour known as negative charge transfer (negative-CT) emerges when the covalent interactions between metal and ligand orbitals become stronger as compared to the positive-CT regime and gives rise to ligand-ligand ($p$-$p$) charge fluctuation. In the negative-CT insulating regime, the electronic ground state is characterized by $d^n \underline{L}$ and the charge-fluctuation is $d^n \underline{L} + d^n \underline{L} = d^n+d^n \underline{L}^2$ type \cite{ref20,ref21,ref22,ref23,ref24,ref25,ref26,ref27,ref28,ref29,ref30,ref31}.

The exploration of the different insulating nature related to charge fluctuation has predominantly been focused on 3$d$-based oxide systems, while post-transition metal oxides with $s$ or $p$ valence orbitals have received less attention. Consequently, the understanding of how different $s$-$s$ or $s$-$p$ charge fluctuations compare to $d$-$d$ or $d$-$p$ charge fluctuations remains unexplored in the literature. One example of a post-transition metal oxide that falls into this category is BaBiO$_3$ (BBO), which is known to exhibit strong negative-CT insulating behaviour in the literature \cite{ref32,adts_sumit}. Also, it has been claimed that the electronic ground state of BBO in the cubic phase can be represented as $Bi^{3+} \underline{L}$, signifying the involvement of the Bi$^{3+}$ cation or lone pair (6$s^2$) and ligand orbitals in the charge and bond-disproportionated phase. The charge fluctuation here can be expressed as Bi$^{3+}\underline{L} + Bi^{3+}\underline{L} = Bi^{3+}+Bi^{3+}\underline{L}^2$ in the negative-CT region, commonly referred to as $p$-$p$ charge fluctuation \cite{ref29,ref30}. It is worth noting that there are also reports suggesting that BaBiO$_3$ is a Peierls insulator \cite{ref33,ref34,ref35,ref36}.

The crystal structure of BBO affects the nature of CD and it exhibits different phases at different temperature ranges. From 4K to 140K, it adopts a monoclinic structure with the space group $P2_1/n$. In the temperature range of 140K to 430K, the structure remains monoclinic but with the space group $I2/m$. Between 430K and 820K, a rhombohedral phase with the space group $R\overline{3}$ is observed. Above 820K, BBO stabilizes in a cubic phase with the space group $Fm\overline{3}m$. These temperature dependent structural transitions are driven by intrinsic octahedral tilting distortion present in BBO \cite{ref37,ref38,ref39}. But apart from the crystal symmetry there have been no prior investigations specifically addressing the relationship between temperature induced variations in charge transfer energy and the resulting alterations in charge disproportionation (CD) at the bismuth (Bi) site and oxygen (O) site. There are some related studies in rare earth nickelates which lies in the negative-CT regime. For example photoelectron spectroscopy investigations of the EuNiO$_3$ in negative-CT regime have uncovered a notable variation in the Ni 2$p$ multiplets with decreasing temperature. Remarkably, distinct indications of Ni 2$p$ multiplet splitting are observed not only below the anticipated metal-insulator transition (MIT), but also albeit to a lesser degree, above the MIT. This peculiar electronic irregularity has been attributed to the occurrence of the dynamic nature of CD, which is associated with the distortion of the NiO$_6$ octahedra within the perovskite structure \cite{ref40}. Similarly, Tajima and coworkers have shown that the Bi 6$s$ bands are split due to a dynamic local CD with a pseudo-gap even in the metallic state of Pb-doped BBO \cite{ref41}. The hypothesis of dynamic charge fluctuations associated with a phononic vibrational amplitude that varies with temperature has been the subject of intense debate. In this context, in bulk BBO, we have previously shown the existence of dynamic CD using temperature-dependent photoelectron spectroscopy studies \cite{ref32}. This CD at the Bi site has a strong implication on the O 2$p$ hole density, which can affect the O 1$s$ core level shape.

In this study, our primary objective is to gain insights into the intricate relationship between covalent and ionic interactions within the Bi-O bond, which ultimately influences the dynamic CD of BBO and the shape of the O 1$s$ core level spectra. To achieve this, we examine the electronic structure of BBO thin film, employing temperature as an external stimulus, an aspect that has not been explored in previous studies. By altering the density of ligand holes using substrate-induced strain, we aim to understand how these interactions modify the dynamic CD of BBO and its impact on the O 1$s$ core level spectra. Utilizing the x-ray photoemission spectroscopy (XPS) valence band spectroscopy (VBS) and resonance photoemission spectra (RPES), we gain insights into the nature of the orbitals near the E$_\text{F}$ and their contributions to the electronic properties. At 300K, we observe that BBO exhibits strong covalent insulating behaviour. This can be attributed to the presence of oxygen 2$p$ orbitals near the E$_\text{F}$. However, as the temperature is decreased, we observe that the system shifts towards an ionic limit. The temperature-dependent changes in the electronic structure of BBO, as revealed by our VBS and RPES, provide valuable insights into the evolution of charge fluctuations and the nature of the insulating behaviour in this material. This also unravels the temperature-induced changes in the metal-ligand hybridization and ligand hole density, providing a comprehensive understanding of the charge state dynamics at the Bi site and ligand site in BBO. This investigation is crucial since the modification of holes in the ligand band can affect the catalytic and superconductivity properties of Bi-based oxide systems.

\section{Methods and Results}

\subsection{Crystal structure}
The thin film preparation of BBO has been carried out on a (110) single crystal substrate of NdGaO$_3$ (NGO) using pulsed laser deposition, which will be referred to as BBON throughout the manuscript. The deposition parameters were followed as described in detail in our previous work [32]. Figure 1(a) illustrates reflection high-energy electron diffraction (RHEED) oscillations during the growth of the film. The sinusoidal type oscillation confirms the layer-by-layer growth of the BBON film. The inset at the bottom of Fig. 1(a) displays the RHEED pattern associated with this observation. The top inset illustrates the intensity profile in the in-plane direction. Notably, the positions of the two consecutive peaks relative to the central peak exhibit symmetry. These findings provide strong evidence for the epitaxial quality of the BBON. The $\theta$-2$\theta$ X-ray diffraction pattern (XRD) of BBON (Fig. 1(b)) indicates a possible pseudo-cubic phase of BBON. The presence of a split in the peak at 2$\theta$ value of 41.5$^\circ$ suggests that the film deviates from its cubic crystal structure. This is because the substrate-induced strain can stabilize different crystal symmetries and also modify the octahedral tilting distortion \cite{ref42,ref43}. The peak split is quite different from the bulk counterpart, indicating reduced tilting angle between two alternate octahedra compared to the bulk monoclinic phase ($I2/m$). Consequently, the film stabilizes in a distorted pseudo-cubic phase.

\begin{figure}[t]
\centering
\includegraphics[
  width=9.0cm,
  height=6.2cm,
  keepaspectratio
]{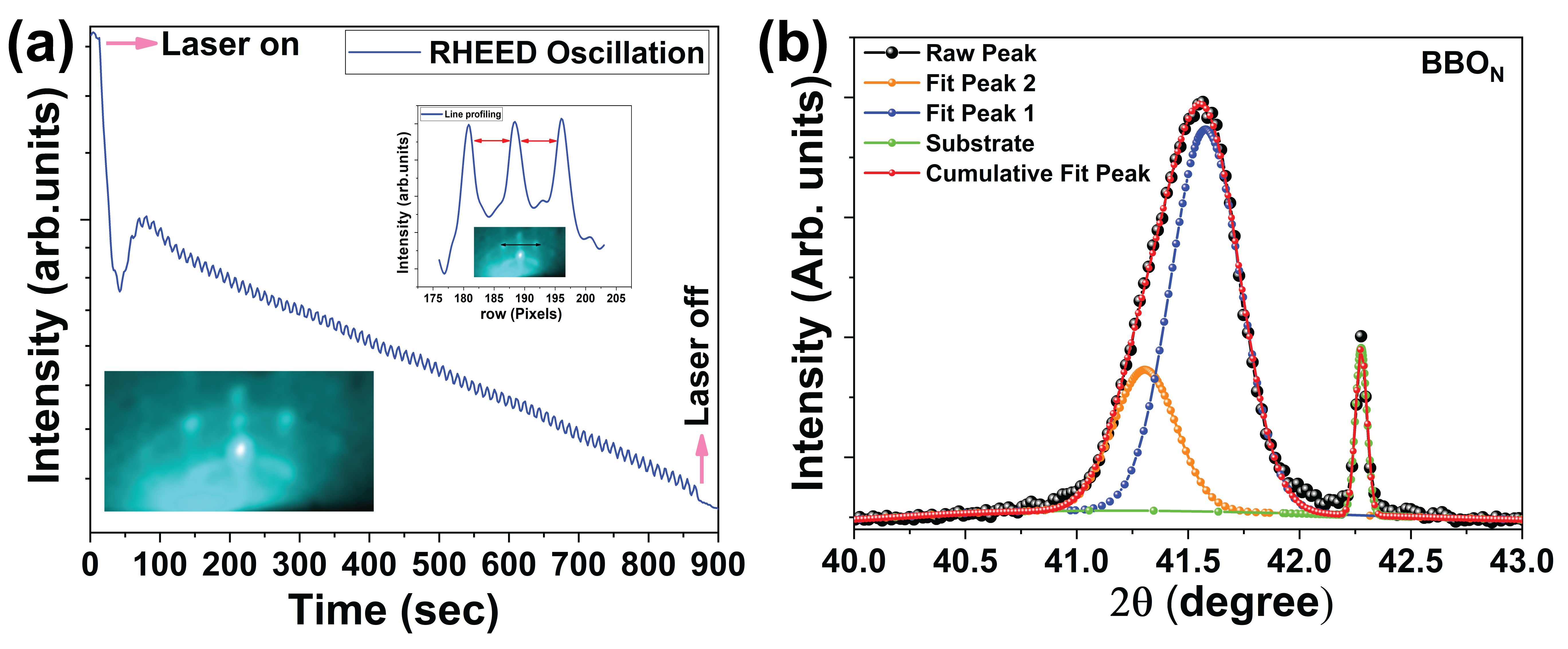}
\caption{\label{fig:fig1}
(a) RHEED oscillations and reciprocal space image with line profile (inset).
(b) Out-of-plane $\theta$--2$\theta$ fitted scan of (002) reflection of BBON thin film.}
\end{figure}

\subsection{Valence band spectra and O 2$p$ resonance photoemission spectra}
To investigate the electronic structure of BBON near the E$_\text{F}$ at different temperatures, we have recorded the valence band spectra using a synchrotron light source (Indus I, RRCAT, Indore, India) by varying the photon energy, which is also known as resonant photoemission spectroscopy (RPES). Figure 2(a)-(d) presents a comparison of the VBS of BBON at 300 K and 160 K for photon energies ranging from 16 eV to 19 eV. These spectra have been normalized well below the E$_\text{F}$ as the shallow core-level remains unchanged with varying photon energy. In Figure 2(a), at 16 eV photon energy, the observed features from E$_\text{F}$ down to -10.0 eV are shown for both 300K and 160K temperatures. Using density functional theory (DFT) calculated partial density of states (PDoS) and VBS in our previous study \cite{ref32}, we showed that feature `a' centred around -3.8 eV arises due to Bi 6$s$/6$p$ - O 2$p$ hybridized state, while the feature `c' in the energy range -6 eV to -8.0 eV corresponds to the Bi 6$p$ - O 2$p$ hybridized states and feature `b' around -5.0 eV is associated with non-bonded oxygen states. 
We aim to investigate the O 2$s$-2$p$ resonance to understand the temperature dependent alteration of the O 2$p$ density of states, which is an outcome of the modification of hole density in the O 2$p$ band. The O 2$s$-2$p$ resonance in O 2$p$ band with hole (O $2p^5\underline{L}$; where $\underline{L}$ is the ligand hole) usually occurs around 19 eV due to the interference between the direct O 2$s$ to O 2$p$ excitation $2s^22p^5\underline{L} \rightarrow 2s^22p^4\underline{L}^2+e^-$ and resonant excitation through meta-stable state $(2s^12p^6)^*$ \cite{ref32,ref44}. At low photon energies, the photoionization cross-section (PIC) of Bi 6$s$ is negligible, but that of Bi 6$p$ and O 2$p$ is relatively higher [32]. Below 19 eV, the PIC of Bi 6$p$ increases. In fig. 2(a) at 16 eV photon energy, the valence band spectra at 300K and 160K appear similar near E$_\text{F}$. The resemblance in the vicinity of the E$_\text{F}$ between the spectra obtained at these two temperatures discards the possibility of surface alterations or contamination occurring during the cooling and heating process. However, a feature centred around -7.7 eV indicated by the arrow in Fig. 2(a) is more pronounced at 160K compared to 300K. This could be due to the increased dominance of Bi 6$p$ orbitals at 160K, especially since at this photon energy, the PIC of Bi 6$p$ is higher than that of O 2$p$. This feature decreases as we move to 17 eV photon energy for the 160K spectra shown in Fig. 2(b). Above 17 eV, from 18 eV photon energy there is an evident increase in spectral weight from E$_\text{F}$ down to -6 eV for 300K compared to 160K, which can be attributed to O 2$p$ orbitals due to their increased PIC with photon energy. Simultaneously, a new feature appears at 19 eV, centered around -4.6 eV for the 300K spectrum. This feature indicates the presence of an O 2$p$ band holes, detected through the O 2$s$-2$p$ resonance process discussed above, suggesting the possible electronic ground state of BBON to be Bi$^{3+}\underline{L}$ at 300K. However, this specific VBS structure is absent in the 160K spectra, indicating a reduction in the hole density in the O 2$p$ band at lower temperatures and potentially reduced hybridization between Bi 6$s$ and O 2$p$. This modification of the hole density with decreasing temperature further affects the CD at the Bi site. In the following section, we will delve into the changes in the charge state of Bi and O by examining the core-level spectra.

\begin{figure}[H] 
\centering
\includegraphics[width=\columnwidth]{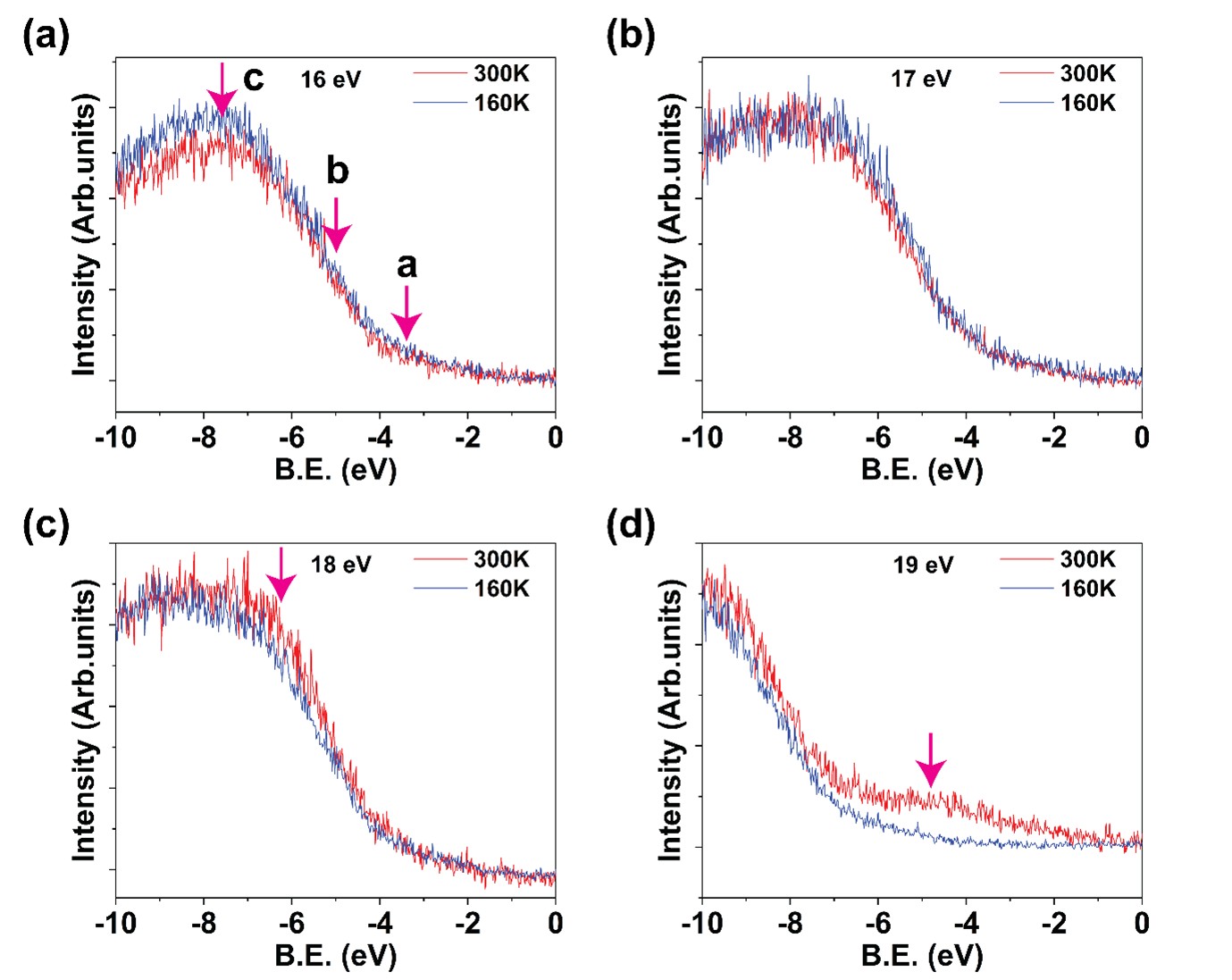} 
\caption{\label{fig:fig2}Comparison of valence band spectra at O 2$s$-2$p$ resonance energy region of BBON thin film at 300K and 160K at (a) 16 eV, (b) 17 eV, (c) 18 eV, and (d) 19 eV photon energy.}
\end{figure}

\subsection{X-ray photoemission spectroscopy}

\subsubsection{Bi 4$f$ core level spectra}
To gain further insights into the charge state variation and modification of the Bi site CD due to the change of electronic density of states near E$_\text{F}$ in the system, XPS measurements were conducted at 300 K and 160 K by examining the Bi 4$f$ core-level spectra. In an ideal Bi$^{4+}$ state, two spin-orbit split peaks, namely 4$f_{7/2}$ and 4$f_{5/2}$, would be observed. However, in the spectrum recorded at 300K, four peaks were observed, as shown in Fig. 3(a). Two intense peaks, labelled as S1 and S2, were observed at higher binding energy positions. The binding energy positions for S1 and S2 were measured to be 157.0 eV and 162.3 eV, respectively, for the 4$f_{7/2}$ and 4$f_{5/2}$ states. The energy difference between S1 and S2 corresponds to the expected spin-orbit splitting of 5.3 eV, which is consistent with previous reports on the Bi 4$f$ core-level spectra \cite{ref32}. In addition to the intense peaks, two shoulder-like features, labelled as R1 and R2, were observed at lower binding energy positions for both the 4$f_{7/2}$ and 4$f_{5/2}$ states. The binding energy positions for R1 and R2 were measured to be 155.1 eV and 160.4 eV, respectively. Notably, the spin-orbit splitting between R1 and R2 was also found to be 5.3 eV. The observation of four distinct peaks with the same spin-orbit splitting suggests the presence of two different charge states, viz. $4+\delta$ and $4-\delta$ at bismuth sites. Figure 3(b) displays the XPS spectra of the Bi 4$f$ core level recorded at 160K. Similar to the spectra obtained at 300K, the presence of shoulder peaks R1 and R2, as well as the intense peaks S1 and S2, was observed. The binding energy positions of S1 and S2 at 160K were measured to be 157.8 eV and 163.2 eV, respectively, while R1 and R2 were found to be located at 155.9 eV and 161.2 eV, respectively. The spin-orbit splitting gaps between S1-S2 and R1-R2 were both determined to be 5.3 eV, consistent with the observations in the 300K spectra. Interestingly, at 160K, the intensity of the R1 and R2 peaks exhibited a significant increase compared to the 300K core-level spectra. Additionally, the full-width at half-maximum (FWHM) of the R1 and R2 peaks also increased relative to the 300K spectra, as indicated in Table \ref{tab:table1}. Moreover, the splitting between R1 and S1, as well as between R2 and S2 slightly increased to 2.1 eV at 160 K as compared to 1.9 eV observed at 300 K. These findings suggest an increase in the $\delta$ value as the temperature decreases. Indeed, the observed enhanced intensity and broadening of the Bi 4$f$ core level peaks at 160K compared to 300K suggest a temperature-dependent modification of the CD at the Bi site in the BBON thin film.
\begin{table}[H] 
\caption{\label{tab:table1} FWHM of Bi 4$f$ core level spectra for 300K and 160K respectively.}
\begin{ruledtabular}
\begin{tabular}{cccccc}
Temperature (K) & \multicolumn{2}{c}{4$f_{7/2}$} & \multicolumn{2}{c}{4$f_{5/2}$} \\
\cline{2-3}\cline{4-5}
& R1 & S1 & R2 & S2 \\
\hline
300 & 1.06 & 1.70 & 0.97 & 1.68 \\
160 & 1.35 & 1.86 & 1.51 & 1.78 \\
\end{tabular}
\end{ruledtabular}
\end{table}

\begin{figure}[H] 
\centering
\includegraphics[width=\columnwidth]{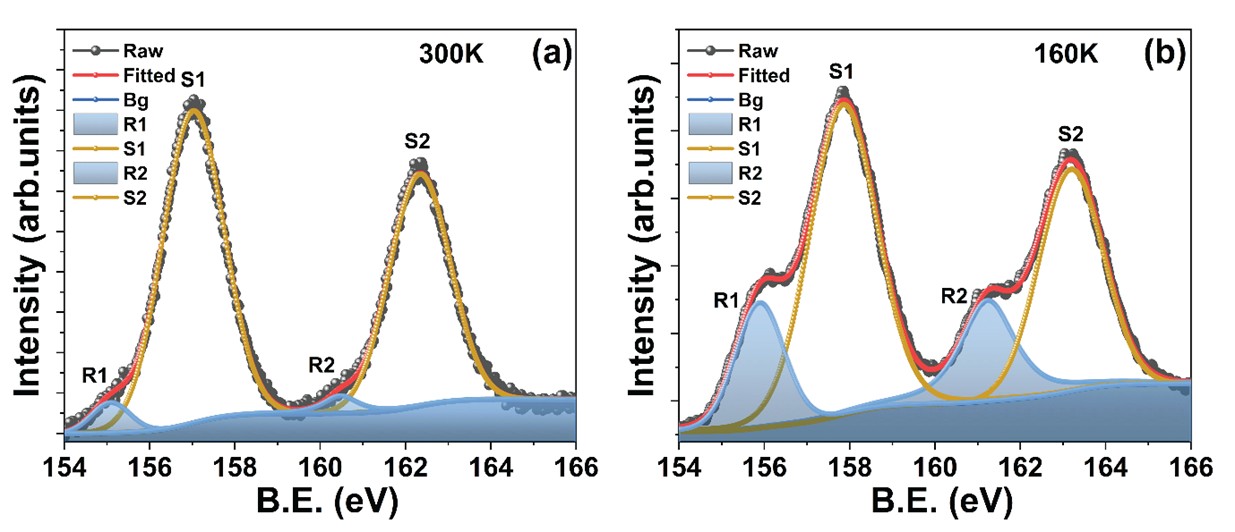} 
\caption{\label{fig:fig3}X-ray photoemission spectra of Bi 4$f$ core-level of BBON thin film at (a) 300K and (b) 160K.}
\end{figure}

\subsubsection{O 1$s$ core level spectra}
The observed temperature-dependent modification of the charge state at the Bi site discussed in the above section also has a strong impact on the hole density distribution in the O 2$p$ band in BiO$_2$ lattice, as observed in RPES results. To further investigate the effect of temperature on the O 2$p$ band hole density, O 1$s$ core level spectra were recorded at 300 K and 160 K as shown in fig. 4(a) and (b). The O 1$s$ spectrum at 300K exhibited an asymmetric behaviour with a shoulder-like feature at the higher binding energy position. A previous work has shown that an asymmetric shape in the O 1$s$ core level can be caused by the existence of O 2$p$ holes \cite{ref45}. Recently, it has been demonstrated that in the case of NdNiO$_3$ and NdNiO$_2$, the O 1$s$ core level exhibits an additional feature at a lower binding energy in the negative-CT insulating regime. The origin of this lower binding energy feature has been attributed to the presence of holes in the Ni-O sublattices \cite{ref45}. The intensity of this characteristic relies on the strength of the core-hole potential or the interaction between the valence electron and the created core-hole in the excited state. If there is no valence hole present in the O 2$p$ band, the core-hole will be well screened from the photoelectron by the valence electron. However, when an O 2$p$ valence hole exists, the screening will be less effective compared to the previous scenario. Consequently, the shoulder-like feature at lower binding energy is observed for the later case. Sawatzky et al. comprehensively demonstrated the variation of hole density in the O 2$p$ band with the changes in the core-hole potential in BBO \cite{ref46} using configuration interaction calculation. The cluster calculation comprises an octahedral structure composed of six oxygen atoms. Each of these oxygen atoms possesses three orbitals (p$_x$, p$_y$ and p$_z$) alongside a core 1$s$ orbital. They also illustrated the influence of Bi-O hybridization on the shape of the O 1$s$ core level spectra. When a core hole is created, weak satellite-like features can be observed in the O 1$s$ spectrum, and the intensity of these satellite peaks and separation from the main peak increase with the strength of the core-hole potential.
We fitted the O 1$s$ core level spectra using three Gaussian peaks, where the two main peaks, O1 and O2, were located at 529.0 eV and 530.6 eV, respectively. These peaks can be understood by considering two excited states $|O_1\rangle$, and $|O_2\rangle$, with $|O_1\rangle$ representing the $O1s^12p^5$ state and $|O_2\rangle$ representing the $O1s^12p^6$ state both of which consists of a single O 1$s$ core hole, and their contributions to the intensity can be expressed as a linear combination of the intensity of the above two states.
\begin{equation}
I_0=\alpha_1I_1+ \alpha_2I_2 \label{eq:eq1}
\end{equation}
Here, $I_1$ and $I_2$ are the intensity corresponding to the $|O_1\rangle$, and $|O_2\rangle$ excited state. The intensity of the O 1$s$ core-level spectra $I_0$, is influenced by the probability amplitudes $\alpha_1$ and $\alpha_2$ of the corresponding states without and with an O 2$p$ valence hole, respectively. From the O 1$s$ core-level spectra at 300K, it is evident that the intensity as well as FWHM of the O1 peak is greater than that of O2, as shown in fig. 4(a) implying that $\alpha_1>\alpha_2$. Whereas at 160K, the intensity of the O2 peak is increased compared to O1, as depicted in fig 4(b), resulting in $\alpha_1<\alpha_2$. However, according to the fitting results, the FWHM values of the two peaks become nearly identical at 160K. The observed FWHM values are given in Table \ref{tab:table2}. These observations further corroborate the RPES results, which revealed a higher O 2$p$ hole density of states near E$_\text{F}$ at 300K as compared to that at 160K and divulged deviations in the charge transfer energy and covalency. It is also noted that the separation between the O1 and O2 peaks has also increased from 1.6 eV to 1.9 eV, which is concomitant with the observed temperature-dependent modifications in the Bi site CD obtained from Bi core level spectra.

\begin{table}[H] 
\caption{\label{tab:table2}FWHM of O 1$s$ core level spectra for 300K and 160K, respectively.}
\begin{ruledtabular}
\begin{tabular}{ccc}
Temperature (K) & \multicolumn{2}{c}{FWHM of O 1$s$} \\
\cline{2-3}
& O1 & O2 \\
\hline
300 & 2.67 & 1.70 \\
160 & 2.10 & 2.11 \\
\end{tabular}
\end{ruledtabular}
\end{table}

\begin{figure}[H] 
\centering
\includegraphics[width=\columnwidth]{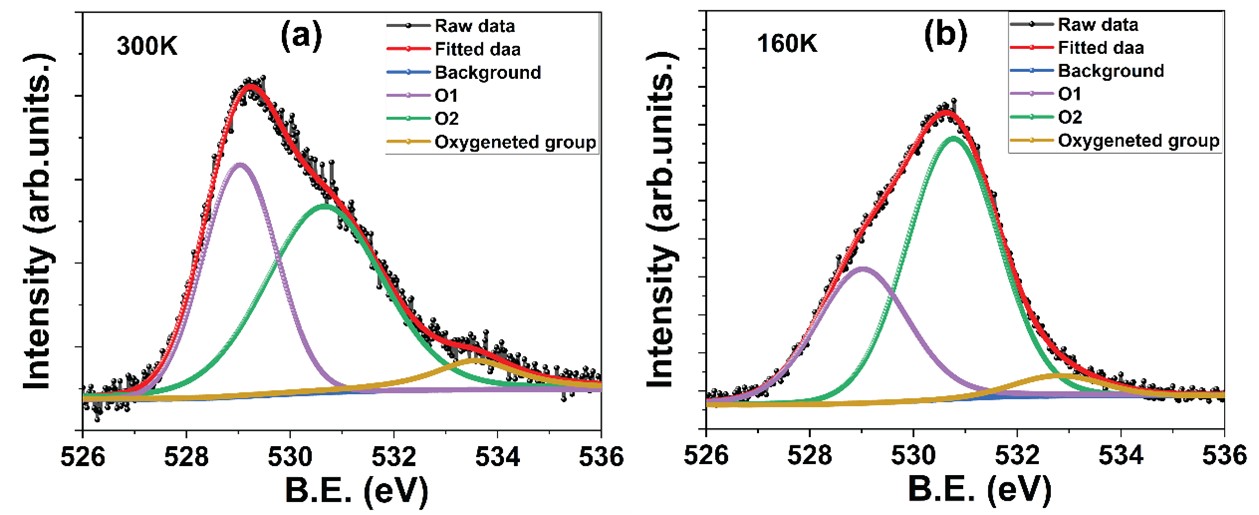} 
\caption{\label{fig:fig4}X-ray photoemission of O 1$s$ core-level spectra of BBON thin film at (left) 300K and (right) 160K.}
\end{figure}

\section{Discussion}
The temperature-dependent variation in the VBS near the E$_\text{F}$ reveals alterations in the hybridization between Bi and O. Specifically, when examining the VBS results at 19 eV, a noticeable resonance is observed at 300K compared to 160K. This indicates a decrease in the Bi-O hybridization as the temperature decreases to 160K, which could be linked to dynamic CD characteristics. The amplitude of dynamic charge fluctuations between $4+\delta$ and $4-\delta$ decreases as temperature decreases, primarily due to the octahedral breathing mode, which ultimately decreases the Bi-O covalent nature and shifts towards an ionic nature. Another possibility is that the temperature-induced alterations affect the tilting of the octahedra. These changes in these two critical parameters result in shifts in charge states or $\delta$ values, as well as changes in the FWHM of the Bi 4$f$ core level spectra when transitioning from 300K to 160K.
Moreover, the O 1$s$ core-level spectra, which are influenced by the O 2$p$ hole density and Bi-O hybridization, exhibit temperature related modifications. At 300K, the spectra display an asymmetric shape with shoulder-like features, indicating the presence of O 2$p$ hole density. Conversely, at 160K, the decreased O 2$p$ hole density is reflected in the increased intensity of the O2 peak and a narrower FWHM, while the O1 peak shows the opposite trend. This suggests that the core-hole potential also changes with temperature, and the screening effect differs at these two temperatures. These findings further support the dynamic nature of CD, as previously observed in bulk BBO \cite{ref32}. Also, this observation indicates the shift towards ionic type from the strong covalent character of the Bi-O bond.

\section{Conclusion}
Our investigations of BBON’s electronic structure, utilizing synchrotron-based VBS, RPES, and lab-based XPS, provide valuable insights into how the CD at Bi site changes as a function of temperature. We observe a strong correlation between the CD at the Bi site and the corresponding density of O 2$p$ holes, which is supported by the observed changes in O 2$p$ hole density as temperature varies. This observation strongly suggests a transition from the predominantly covalent nature of the Bi-O bond to a more pronounced ionic character. This phenomenon of CD modification with temperature has been previously noted in materials like EuNiO$_3$ which is known to be negative-CT insulator; however, such behaviour for post-transition metal oxide like BBO is rather new. Our findings regarding temperature-dependent modifications in electronic structure, CD at Bi site and ligand hole density offer valuable insights into the covalent and ionic behavior of BBON. These results underscore the significance of metal-ligand hybridization and temperature-induced variations in the electronic properties of the system, which has strong relevance to catalytic and superconductive properties in bismuth oxides and related systems.

\section{Acknowledgments}
Mr. Sharad Karwal and Mr. A. Wadekar are acknowledged for their assistance during the temperature-dependent XPS and VBS measurements carried out at Indus-I Synchrotron, BL-2, Raja Ramanna Centre for Advanced Technology, Indore, India.

\bibliographystyle{apsrev4-2}
\bibliography{bbo}

@article{ref1,
  title = {Metal-insulator transitions},
  author = {Imada, Masatoshi and Fujimori, Atsushi and Tokura, Yoshinori},
  journal = {Rev. Mod. Phys.},
  volume = {70},
  issue = {4},
  pages = {1039--1263},
  numpages = {0},
  year = {1998},
  month = {Oct},
  publisher = {American Physical Society},
  doi = {10.1103/RevModPhys.70.1039},
  url = {https://link.aps.org/doi/10.1103/RevModPhys.70.1039}
}

@article{ref2,
  title = {Metal-Insulator Transition},
  author = {Mott, N. F.},
  journal = {Rev. Mod. Phys.},
  volume = {40},
  issue = {4},
  pages = {677--683},
  numpages = {0},
  year = {1968},
  month = {Oct},
  publisher = {American Physical Society},
  doi = {10.1103/RevModPhys.40.677},
  url = {https://link.aps.org/doi/10.1103/RevModPhys.40.677}
}

@article{ref3,
  title = {Doping a Mott insulator: Physics of high-temperature superconductivity},
  author = {Lee, Patrick A. and Nagaosa, Naoto and Wen, Xiao-Gang},
  journal = {Rev. Mod. Phys.},
  volume = {78},
  issue = {1},
  pages = {17--85},
  numpages = {0},
  year = {2006},
  month = {Jan},
  publisher = {American Physical Society},
  doi = {10.1103/RevModPhys.78.17},
  url = {https://link.aps.org/doi/10.1103/RevModPhys.78.17}
}

@article{ref4,
  title = {The Anderson-Mott transition},
  author = {Belitz, D. and Kirkpatrick, T. R.},
  journal = {Rev. Mod. Phys.},
  volume = {66},
  issue = {2},
  pages = {261--380},
  numpages = {0},
  year = {1994},
  month = {Apr},
  publisher = {American Physical Society},
  doi = {10.1103/RevModPhys.66.261},
  url = {https://link.aps.org/doi/10.1103/RevModPhys.66.261}
}

@book{ref5AshcroftMermin,
  author    = {Ashcroft, N. W. and Mermin, N. D.},
  title     = {Solid State Physics},
  publisher = {Saunders College},
  address   = {Philadelphia, PA},
  year      = {1976}
}

@book{ref6,
author = {Roy, Sindhunil Barman},
title = {Mott Insulators},
publisher = {IOP Publishing},
year = {2019},
series = {2053-2563},
isbn = {978-0-7503-1596-8},
url = {https://dx.doi.org/10.1088/2053-2563/ab16c9},
doi = {10.1088/2053-2563/ab16c9}
}

@book{ref7,
author = {Aylett, B. J.},
title = {Transition Metal Oxides: Structure, Properties and Synthesis of Ceramic Oxides. C.N.R. Rao and B. Raveau. 2nd edn. Wiley–VCH, New York and Weinheim, 1998.},
publisher = {Wiley-VCH},
journal = {Applied Organometallic Chemistry},
volume = {13},
pages = {476-477},
doi = {https://doi.org/10.1002/(SICI)1099-0739(199906)13:6<476::AID-AOC851>3.0.CO;2-N},
year = {1999}
}

@article{ref8,
  title = {Band gaps and electronic structure of transition-metal compounds},
  author = {Zaanen, J. and Sawatzky, G. A. and Allen, J. W.},
  journal = {Phys. Rev. Lett.},
  volume = {55},
  issue = {4},
  pages = {418--421},
  numpages = {0},
  year = {1985},
  month = {Jul},
  publisher = {American Physical Society},
  doi = {10.1103/PhysRevLett.55.418},
  url = {https://link.aps.org/doi/10.1103/PhysRevLett.55.418}
}

@article{ref9,
  title = {Mean-field results of the multiple-band extended Hubbard model for the square-planar ${\mathrm{CuO}}_{2}$ lattice},
  author = {Nimkar, Seva and Sarma, D. D. and Krishnamurthy, H. R. and Ramasesha, S.},
  journal = {Phys. Rev. B},
  volume = {48},
  issue = {10},
  pages = {7355--7363},
  numpages = {0},
  year = {1993},
  month = {Sep},
  publisher = {American Physical Society},
  doi = {10.1103/PhysRevB.48.7355},
  url = {https://link.aps.org/doi/10.1103/PhysRevB.48.7355}
}

@article{ref10,
  title = {Tuning the bond-order wave phase in the half-filled extended Hubbard model},
  author = {Kumar, Manoranjan and Ramasesha, S. and Soos, Z. G.},
  journal = {Phys. Rev. B},
  volume = {79},
  issue = {3},
  pages = {035102},
  numpages = {8},
  year = {2009},
  month = {Jan},
  publisher = {American Physical Society},
  doi = {10.1103/PhysRevB.79.035102},
  url = {https://link.aps.org/doi/10.1103/PhysRevB.79.035102}
}

@article{ref11,
  title = {Intersite Coulomb Interactions in Charge-Ordered Systems},
  author = {Jang, Bo Gyu and Kim, Minjae and Lee, Sang-Hoon and Yang, Wooil and Jhi, Seung-Hoon and Son, Young-Woo},
  journal = {Phys. Rev. Lett.},
  volume = {130},
  issue = {13},
  pages = {136401},
  numpages = {7},
  year = {2023},
  month = {Mar},
  publisher = {American Physical Society},
  doi = {10.1103/PhysRevLett.130.136401},
  url = {https://link.aps.org/doi/10.1103/PhysRevLett.130.136401}
}

@article{ref12,
  title = {Strength of the Hubbard potential and its modification by breathing distortion in ${\mathrm{BaBiO}}_{3}$},
  author = {Lukyanov, Alexander E. and Kovalev, Ivan A. and Neverov, Vyacheslav D. and Zhumagulov, Yaroslav V. and Krasavin, Andrey V. and Kochan, Denis},
  journal = {Phys. Rev. B},
  volume = {105},
  issue = {4},
  pages = {045131},
  numpages = {10},
  year = {2022},
  month = {Jan},
  publisher = {American Physical Society},
  doi = {10.1103/PhysRevB.105.045131},
  url = {https://link.aps.org/doi/10.1103/PhysRevB.105.045131}
}

@article{ref13,
  title = {Mott-Hubbard type insulating nature of epitaxial $\mathrm{LaV}{\mathrm{O}}_{3}$ thin films},
  author = {Jana, Anupam and Choudhary, R. J. and Phase, D. M.},
  journal = {Phys. Rev. B},
  volume = {98},
  issue = {7},
  pages = {075124},
  numpages = {10},
  year = {2018},
  month = {Aug},
  publisher = {American Physical Society},
  doi = {10.1103/PhysRevB.98.075124},
  url = {https://link.aps.org/doi/10.1103/PhysRevB.98.075124}
}

@article{ref14,
  title = {Insulating phases of vanadium dioxide are Mott-Hubbard insulators},
  author = {Huffman, T. J. and Hendriks, C. and Walter, E. J. and Yoon, Joonseok and Ju, Honglyoul and Smith, R. and Carr, G. L. and Krakauer, H. and Qazilbash, M. M.},
  journal = {Phys. Rev. B},
  volume = {95},
  issue = {7},
  pages = {075125},
  numpages = {6},
  year = {2017},
  month = {Feb},
  publisher = {American Physical Society},
  doi = {10.1103/PhysRevB.95.075125},
  url = {https://link.aps.org/doi/10.1103/PhysRevB.95.075125}
}

@article{ref15,
doi = {10.1088/0953-8984/21/32/323202},
url = {https://dx.doi.org/10.1088/0953-8984/21/32/323202},
year = {2009},
month = {jul},
publisher = {},
volume = {21},
number = {32},
pages = {323202},
author = {Perucchi, A and Baldassarre, L and Postorino, P and Lupi, S},
title = {Optical properties across the insulator to metal transitions in vanadium oxide
compounds},
journal = {J. Phys. Condens. Matter}
}

@article{ref16,
  title = {Enhanced charge-transfer character in the monoclinic phase of Mott insulator $\mathrm{La}\mathrm{V}{\mathrm{O}}_{3}$ thin films},
  author = {Jana, Anupam and Raghunathan, Rajamani and Rawat, Ritu and Choudhary, R. J. and Phase, D. M.},
  journal = {Phys. Rev. B},
  volume = {102},
  issue = {23},
  pages = {235108},
  numpages = {11},
  year = {2020},
  month = {Dec},
  publisher = {American Physical Society},
  doi = {10.1103/PhysRevB.102.235108},
  url = {https://link.aps.org/doi/10.1103/PhysRevB.102.235108}
}

@article{ref17,
author = {Chowdhury, Sourav and Jana, Anupam and Kuila, Manik and Reddy, Varimalla R. and Choudhary, Ram J. and Phase, Deodatta M.},
title = {Negative Charge-Transfer Energy in SrCoO2.5 Thin Films: An Interplay between O-2p Hole Density, Charge-Transfer Energy, Charge Disproportionation, and Ferromagnetic Ordering},
journal = {ACS Appl. Electron. Mater},
volume = {2},
number = {12},
pages = {3859-3870},
year = {2020},
doi = {10.1021/acsaelm.0c00698},
}

@article{ref18,
  title = {Anomalous Electronic State in ${\mathrm{CaCrO}}_{3}$ and ${\mathrm{SrCrO}}_{3}$},
  author = {Zhou, J.-S. and Jin, C.-Q. and Long, Y.-W. and Yang, L.-X. and Goodenough, J. B.},
  journal = {Phys. Rev. Lett.},
  volume = {96},
  issue = {4},
  pages = {046408},
  numpages = {4},
  year = {2006},
  month = {Feb},
  publisher = {American Physical Society},
  doi = {10.1103/PhysRevLett.96.046408},
  url = {https://link.aps.org/doi/10.1103/PhysRevLett.96.046408}
}

@article{ref19,
  title = {Hubbard band versus oxygen vacancy states in the correlated electron metal ${\mathrm{SrVO}}_{3}$},
  author = {Backes, S. and R\"odel, T. C. and Fortuna, F. and Frantzeskakis, E. and Le F\`evre, P. and Bertran, F. and Kobayashi, M. and Yukawa, R. and Mitsuhashi, T. and Kitamura, M. and Horiba, K. and Kumigashira, H. and Saint-Martin, R. and Fouchet, A. and Berini, B. and Dumont, Y. and Kim, A. J. and Lechermann, F. and Jeschke, H. O. and Rozenberg, M. J. and Valent\'{\i}, R. and Santander-Syro, A. F.},
  journal = {Phys. Rev. B},
  volume = {94},
  issue = {24},
  pages = {241110},
  numpages = {7},
  year = {2016},
  month = {Dec},
  publisher = {American Physical Society},
  doi = {10.1103/PhysRevB.94.241110},
  url = {https://link.aps.org/doi/10.1103/PhysRevB.94.241110}
}

@article{ref20,
  title = {Covalency-driven unusual metal-insulator transition in nickelates},
  author = {Barman, S. R. and Chainani, A. and Sarma, D. D.},
  journal = {Phys. Rev. B},
  volume = {49},
  issue = {12},
  pages = {8475--8478},
  numpages = {0},
  year = {1994},
  month = {Mar},
  publisher = {American Physical Society},
  doi = {10.1103/PhysRevB.49.8475},
  url = {https://link.aps.org/doi/10.1103/PhysRevB.49.8475}
}

@article{ref21,
  title = {Origin of the band gap in the negative charge-transfer-energy compound ${\mathrm{NaCuO}}_{2}$},
  author = {Mizokawa, T. and Namatame, H. and Fujimori, A. and Akeyama, K. and Kondoh, H. and Kuroda, H. and Kosugi, N.},
  journal = {Phys. Rev. Lett.},
  volume = {67},
  issue = {12},
  pages = {1638--1641},
  numpages = {0},
  year = {1991},
  month = {Sep},
  publisher = {American Physical Society},
  doi = {10.1103/PhysRevLett.67.1638},
  url = {https://link.aps.org/doi/10.1103/PhysRevLett.67.1638}
}

@article{ref22,
  author    = {Bisogni, Valentina and Catalano, Sara and Green, Robert J. and Gibert, Marta and Scherwitzl, Raoul and Huang, Yaobo and Strocov, Vladimir N. and Zubko, Pavlo and Balandeh, Shadi and Triscone, Jean-Marc and Sawatzky, George and Schmitt, Thorsten},
  title     = {Ground-state oxygen holes and the metal--insulator transition in the negative charge-transfer rare-earth nickelates},
  journal   = {Nat. Commun.},
  volume    = {7},
  number    = {1},
  pages     = {13017},
  year      = {2016},
  doi       = {10.1038/ncomms13017},
  url       = {https://doi.org/10.1038/ncomms13017},
  issn      = {2041-1723},
}

@article{ref23,
  author    = {Patel, Ranjan Kumar and Patra, Krishnendu and Ojha, Shashank Kumar and Kumar, Siddharth and Sarkar, Sagar and Saha, Akash and Bhattacharya, Nandana and Freeland, John W. and Kim, Jong-Woo and Ryan, Philip J. and Mahadevan, Priya and Middey, Srimanta},
  title     = {Hole doping in a negative charge transfer insulator},
  journal   = {Commun. Phys.},
  volume    = {5},
  number    = {1},
  pages     = {216},
  year      = {2022},
  doi       = {10.1038/s42005-022-00993-1},
  url       = {https://doi.org/10.1038/s42005-022-00993-1},
  issn      = {2399-3650},
}

@article{ref24,
  title = {Theoretical proposal and material realization of ferromagnetic negative charge-transfer energy insulator},
  author = {Liu, Zhao and Li, Xingxing and Zhu, W. and Wang, Z. F. and Yang, Jinlong},
  journal = {Phys. Rev. B},
  volume = {107},
  issue = {1},
  pages = {014413},
  numpages = {22},
  year = {2023},
  month = {Jan},
  publisher = {American Physical Society},
  doi = {10.1103/PhysRevB.107.014413},
  url = {https://link.aps.org/doi/10.1103/PhysRevB.107.014413}
}

@article{ref25,
  title = {Charge Disproportionation without Charge Transfer in the Rare-Earth-Element Nickelates as a Possible Mechanism for the Metal-Insulator Transition},
  author = {Johnston, Steve and Mukherjee, Anamitra and Elfimov, Ilya and Berciu, Mona and Sawatzky, George A.},
  journal = {Phys. Rev. Lett.},
  volume = {112},
  issue = {10},
  pages = {106404},
  numpages = {5},
  year = {2014},
  month = {Mar},
  publisher = {American Physical Society},
  doi = {10.1103/PhysRevLett.112.106404},
  url = {https://link.aps.org/doi/10.1103/PhysRevLett.112.106404}
}

@article{ref26,
  title = {Metal-insulator transition and local-moment collapse in negative charge transfer ${\mathrm{CaFeO}}_{3}$ under pressure},
  author = {Leonov, I.},
  journal = {Phys. Rev. B},
  volume = {105},
  issue = {3},
  pages = {035157},
  numpages = {9},
  year = {2022},
  month = {Jan},
  publisher = {American Physical Society},
  doi = {10.1103/PhysRevB.105.035157},
  url = {https://link.aps.org/doi/10.1103/PhysRevB.105.035157}
}

@article{ref27,
  title = {Electronic structure of negative charge transfer ${\mathrm{CaFeO}}_{3}$ across the metal-insulator transition},
  author = {Rogge, Paul C. and Chandrasena, Ravini U. and Cammarata, Antonio and Green, Robert J. and Shafer, Padraic and Lefler, Benjamin M. and Huon, Amanda and Arab, Arian and Arenholz, Elke and Lee, Ho Nyung and Lee, Tien-Lin and Nem\ifmmode \check{s}\else \v{s}\fi{}\'ak, Slavom\'{\i}r and Rondinelli, James M. and Gray, Alexander X. and May, Steven J.},
  journal = {Phys. Rev. Mater.},
  volume = {2},
  issue = {1},
  pages = {015002},
  numpages = {8},
  year = {2018},
  month = {Jan},
  publisher = {American Physical Society},
  doi = {10.1103/PhysRevMaterials.2.015002},
  url = {https://link.aps.org/doi/10.1103/PhysRevMaterials.2.015002}
}

@article{ref28,
  title = {Spin and charge ordering in self-doped Mott insulators},
  author = {Mizokawa, T. and Khomskii, D. I. and Sawatzky, G. A.},
  journal = {Phys. Rev. B},
  volume = {61},
  issue = {17},
  pages = {11263--11266},
  numpages = {0},
  year = {2000},
  month = {May},
  publisher = {American Physical Society},
  doi = {10.1103/PhysRevB.61.11263},
  url = {https://link.aps.org/doi/10.1103/PhysRevB.61.11263}
}

@article{ref29,
  title = {Oxygen holes and hybridization in the bismuthates},
  author = {Khazraie, Arash and Foyevtsova, Kateryna and Elfimov, Ilya and Sawatzky, George A.},
  journal = {Phys. Rev. B},
  volume = {97},
  issue = {7},
  pages = {075103},
  numpages = {9},
  year = {2018},
  month = {Feb},
  publisher = {American Physical Society},
  doi = {10.1103/PhysRevB.97.075103},
  url = {https://link.aps.org/doi/10.1103/PhysRevB.97.075103}
}

@article{ref30,
  title = {Hybridization effects and bond disproportionation in the bismuth perovskites},
  author = {Foyevtsova, Kateryna and Khazraie, Arash and Elfimov, Ilya and Sawatzky, George A.},
  journal = {Phys. Rev. B},
  volume = {91},
  issue = {12},
  pages = {121114},
  numpages = {5},
  year = {2015},
  month = {Mar},
  publisher = {American Physical Society},
  doi = {10.1103/PhysRevB.91.121114},
  url = {https://link.aps.org/doi/10.1103/PhysRevB.91.121114}
}

@article{ref31,
  title = {Electronic structure of the bond disproportionated bismuthate ${\mathrm{Ag}}_{2}{\mathrm{BiO}}_{3}$},
  author = {Oudah, Mohamed and Kim, Minu and Rabinovich, Ksenia S. and Foyevtsova, Kateryna and McNally, Graham and Kilic, Berkay and K\"uster, Kathrin and Green, Robert and Boris, Alexander V. and Sawatzky, George and Schnyder, Andreas P. and Bonn, D. A. and Keimer, Bernhard and Takagi, Hidenori},
  journal = {Phys. Rev. Mater.},
  volume = {5},
  issue = {6},
  pages = {064202},
  numpages = {9},
  year = {2021},
  month = {Jun},
  publisher = {American Physical Society},
  doi = {10.1103/PhysRevMaterials.5.064202},
  url = {https://link.aps.org/doi/10.1103/PhysRevMaterials.5.064202}
}

@article{ref32,
author = {Sarkar, Sumit and Raghunathan, Rajamani and Chowdhury, Sourav and Choudhary, Ram Janay and Phase, Deodatta Moreshwar},
title = {The Mystery behind Dynamic Charge Disproportionation in BaBiO3},
journal = {Nano Letters},
volume = {21},
number = {19},
pages = {8433-8438},
year = {2021},
doi = {10.1021/acs.nanolett.1c03103},

    
    

}

@article{ref33,
  title = {Polaronic Hole Trapping in Doped ${\mathrm{BaBiO}}_{3}$},
  author = {Franchini, C. and Kresse, G. and Podloucky, R.},
  journal = {Phys. Rev. Lett.},
  volume = {102},
  issue = {25},
  pages = {256402},
  numpages = {4},
  year = {2009},
  month = {Jun},
  publisher = {American Physical Society},
  doi = {10.1103/PhysRevLett.102.256402},
  url = {https://link.aps.org/doi/10.1103/PhysRevLett.102.256402}
}

@article{ref34,
  title = {Missing valence states, diamagnetic insulators, and superconductors},
  author = {Varma, C. M.},
  journal = {Phys. Rev. Lett.},
  volume = {61},
  issue = {23},
  pages = {2713--2716},
  numpages = {0},
  year = {1988},
  month = {Dec},
  publisher = {American Physical Society},
  doi = {10.1103/PhysRevLett.61.2713},
  url = {https://link.aps.org/doi/10.1103/PhysRevLett.61.2713}
}

@article{ref35,
  title = {Madelung energy of the valence-skipping compound $\mathrm{Ba}\mathrm{Bi}{\mathrm{O}}_{3}$},
  author = {Hase, Izumi and Yanagisawa, Takashi},
  journal = {Phys. Rev. B},
  volume = {76},
  issue = {17},
  pages = {174103},
  numpages = {4},
  year = {2007},
  month = {Nov},
  publisher = {American Physical Society},
  doi = {10.1103/PhysRevB.76.174103},
  url = {https://link.aps.org/doi/10.1103/PhysRevB.76.174103}
}

@article{ref36,
  title = {Structural, vibrational, and quasiparticle properties of the Peierls semiconductor ${\text{BaBiO}}_{3}$: A hybrid functional and self-consistent $\text{GW}+\text{vertex-corrections}$ study},
  author = {Franchini, C. and Sanna, A. and Marsman, M. and Kresse, G.},
  journal = {Phys. Rev. B},
  volume = {81},
  issue = {8},
  pages = {085213},
  numpages = {7},
  year = {2010},
  month = {Feb},
  publisher = {American Physical Society},
  doi = {10.1103/PhysRevB.81.085213},
  url = {https://link.aps.org/doi/10.1103/PhysRevB.81.085213}
}

@article{ref37,
title = {Bismuthates: BaBiO3 and related superconducting phases},
journal = {Phys. C Supercond. it Appl.},
volume = {514},
pages = {152-165},
year = {2015},
issn = {0921-4534},
doi = {https://doi.org/10.1016/j.physc.2015.02.012},
url = {https://www.sciencedirect.com/science/article/pii/S0921453415000398},
author = {Arthur W. Sleight},
}

@article{ref38,
title = {Crystal structure of Ba2Bi3+Bi5+O6},
journal = {Solid State Commun.},
volume = {19},
number = {10},
pages = {969-973},
year = {1976},
issn = {0038-1098},
doi = {https://doi.org/10.1016/0038-1098(76)90632-3},
url = {https://www.sciencedirect.com/science/article/pii/0038109876906323},
author = {D.E. Cox and A.W. Sleight}
}

@article{ref39,
doi = {10.35848/1347-4065/abb00d},
url = {https://dx.doi.org/10.35848/1347-4065/abb00d},
year = {2020},
month = {sep},
publisher = {IOP Publishing},
volume = {59},
number = {9},
pages = {095505},
author = {Zhao, Qing and Abe, Tomohiro and Moriyoshi, Chikako and Kim, Sangwook and Taguchi, Ayako and Moriwake, Hiroki and Sun, Hong-Tao and Kuroiwa, Yoshihiro},
title = {Charge order of bismuth ions and nature of chemical bonds in double perovskite-type oxide BaBiO3 visualized by synchrotron radiation X-ray diffraction},
journal = {Jpn. J. Appl. Phys.}
}

@article{ref40,
  title = {Evidence of charge disproportionation on the nickel sublattice in ${\text{EuNiO}}_{3}$ thin films: X-ray photoemission studies},
  author = {Bilewska, K. and Wolna, E. and Edely, M. and Ruello, P. and Szade, J.},
  journal = {Phys. Rev. B},
  volume = {82},
  issue = {16},
  pages = {165105},
  numpages = {5},
  year = {2010},
  month = {Oct},
  publisher = {American Physical Society},
  doi = {10.1103/PhysRevB.82.165105},
  url = {https://link.aps.org/doi/10.1103/PhysRevB.82.165105}
}

@article{ref41,
  title = {Optical study of the metal-semiconductor transition in ${\mathrm{BaPb}}_{1\mathrm{\ensuremath{-}}\mathrm{x}}$${\mathrm{Bi}}_{\mathrm{x}}$${\mathrm{O}}_{3}$},
  author = {Tajima, S. and Uchida, S. and Masaki, A. and Takagi, H. and Kitazawa, K. and Tanaka, S. and Katsui, A.},
  journal = {Phys. Rev. B},
  volume = {32},
  issue = {10},
  pages = {6302--6311},
  numpages = {0},
  year = {1985},
  month = {Nov},
  publisher = {American Physical Society},
  doi = {10.1103/PhysRevB.32.6302},
  url = {https://link.aps.org/doi/10.1103/PhysRevB.32.6302}
}

@article{ref42,
  title = {Partial suppression of structural distortion in epitaxially grown ${\text{BaBiO}}_{3}$ thin films},
  author = {Inumaru, Kei and Miyata, Hajime and Yamanaka, Shoji},
  journal = {Phys. Rev. B},
  volume = {78},
  issue = {13},
  pages = {132507},
  numpages = {4},
  year = {2008},
  month = {Oct},
  publisher = {American Physical Society},
  doi = {10.1103/PhysRevB.78.132507},
  url = {https://link.aps.org/doi/10.1103/PhysRevB.78.132507}
}

@article{ref43,
  title = {Suppression of Three-Dimensional Charge Density Wave Ordering via Thickness Control},
  author = {Kim, Gideok and Neumann, Michael and Kim, Minu and Le, Manh Duc and Kang, Tae Dong and Noh, Tae Won},
  journal = {Phys. Rev. Lett.},
  volume = {115},
  issue = {22},
  pages = {226402},
  numpages = {5},
  year = {2015},
  month = {Nov},
  publisher = {American Physical Society},
  doi = {10.1103/PhysRevLett.115.226402},
  url = {https://link.aps.org/doi/10.1103/PhysRevLett.115.226402}
}

@article{ref44,
  title = {Resonant-photoemission study of ${\mathrm{Ba}}_{0.6}$${\mathrm{K}}_{0.4}$${\mathrm{BiO}}_{3}$},
  author = {Ruckman, M. W. and Di Marzio, D. and Jeon, Y. and Liang, G. and Chen, J. and Croft, M. and Hegde, M. S. and Barboux, P.},
  journal = {Phys. Rev. B},
  volume = {39},
  issue = {10},
  pages = {7359--7362},
  numpages = {0},
  year = {1989},
  month = {Apr},
  publisher = {American Physical Society},
  doi = {10.1103/PhysRevB.39.7359},
  url = {https://link.aps.org/doi/10.1103/PhysRevB.39.7359}
}

@article{ref45,
  author    = {Fu, Ying and Wang, Le and Cheng, Hu and Pei, Shenghai and Zhou, Xuefeng and Chen, Jian and Wang, Shaoheng and Zhao, Ran and Jiang, Wenrui and Liu, Cai and Huang, Mingyuan and Wang, XinWei and Zhao, Yusheng and Yu, Dapeng and Ye, Fei and Wang, Shanmin and Mei, Jia-Wei},
  title     = {Core-level x-ray photoemission and Raman spectroscopy studies on electronic structures in Mott-Hubbard type nickelate oxide NdNiO$_2$},
  journal   = {arXiv preprint arXiv:1911.03177},
  year      = {2019},
  eprint    = {1911.03177},
  archiveprefix = {arXiv},
  primaryclass = {cond-mat.supr-con},
  doi       = {10.48550/arXiv.1911.03177},
  url       = {https://arxiv.org/abs/1911.03177},
}

@article{ref46,
  title = {Experimental and theoretical study of the electronic structure of single-crystal ${\mathrm{BaBiO}}_{3}$},
  author = {Balandeh, Shadi and Green, Robert J. and Foyevtsova, Kateryna and Chi, Shun and Foyevtsov, Oleksandr and Li, Fengmiao and Sawatzky, George A.},
  journal = {Phys. Rev. B},
  volume = {96},
  issue = {16},
  pages = {165127},
  numpages = {7},
  year = {2017},
  month = {Oct},
  publisher = {American Physical Society},
  doi = {10.1103/PhysRevB.96.165127},
  url = {https://link.aps.org/doi/10.1103/PhysRevB.96.165127}
}

@article{adts_sumit,
author = {Sarkar, Sumit and Choudhary, Ram Janay and Sharma, Manju and Raghunathan, Rajamani},
title = {Competing s-p and p-p Fluctuations in Charge-Disproportionation of BaBiO3},
journal = {Adv. Theory Simul.},
volume = {7},
number = {8},
pages = {2400328},
doi = {https://doi.org/10.1002/adts.202400328},
url = {https://advanced.onlinelibrary.wiley.com/doi/abs/10.1002/adts.202400328},
year = {2024}
}

\end{document}